\documentclass[%
reprint,
amsmath,amssymb,
aps,
pre,
]{revtex4-1}

\usepackage{graphicx}
\usepackage{array}
\usepackage{dcolumn}
\usepackage{bm}
\usepackage[utf8]{inputenc}
\usepackage[english]{babel}
\usepackage{amsmath}    
\usepackage{graphicx}   
\usepackage{verbatim}   
\usepackage[usenames, dvipsnames]{color}      
\usepackage{subfigure}  
\usepackage{hyperref}   
\hypersetup{colorlinks=true,citecolor=blue,linkcolor=red,urlcolor=blue}
\usepackage{color}

\definecolor{green}{rgb}{0,0.5,0}

\newcommand*{\dd}{\mathrm{d}}

\newcommand{\B}[1]{{\bm{#1}}}

\begin{document}

\title{Oscillatory Instabilities in 3-Dimensional Frictional Granular Matter}
\author{Silvia Bonfanti$^1$, Joyjit Chattoraj$^2$, Roberto Guerra$^1$, Itamar Procaccia$^3$ and Stefano Zapperi$^{1,4}$}
\affiliation{
$^1$ Center for Complexity and Biosystems, Department of Physics, University of Milan, via Celoria 16, 20133 Milano, Italy;
\\$^2$ School of Physical and Mathematical Sciences,	Nanyang Technological University, Singapore;
\\$^3$ Dept. of Chemical Physics, The Weizmann Institute of Science, Rehovot 76100, Israel.
\\ $^4$ CNR - Consiglio Nazionale delle Ricerche,  Istituto di Chimica della Materia Condensata e di Tecnologie per l'Energia, Via R. Cozzi 53, 20125 Milano, Italy
}
\date{\today}

\begin{abstract}
The dynamics of amorphous granular matter with frictional interactions cannot be derived in general from a Hamiltonian and therefore displays oscillatory instabilities stemming from the onset of complex eigenvalues in the stability matrix. These instabilities were discovered in the context of one and two dimensional systems, while the three dimensional case was never
studied in detail. Here we fill this gap by deriving and demonstrating the presence of oscillatory instabilities in a three dimensional granular packing. We study binary assemblies of spheres of two sizes interacting via classical Hertz and Mindlin force laws for the longitudinal and tangent interactions, respectively.  We formulate analytically the stability matrix in 3D and observe that a couple of complex eigenvalues emerges at the onset of the instability as in the case of frictional disks in two-dimensions. The dynamics then shows oscillatory exponential growth in the Mean-Square-Displacement, followed by a catastrophic event. The generality of these results for any choice of forces that break the symplectic Hamiltonian symmetry is discussed.
\end{abstract}

\maketitle

\section{Introduction}
\label{sec.intro}

The mechanics of dense granular matter has attracted a wide interest for decades as
a paradigmatic example of disordered glassy systems 
\cite{98LN,01OLLN,cates1998jamming,majmudar2007jamming} and for its importance
for technological applications in several fields, from pharmaceutics to agriculture \cite{masuda2006powder}.  In contrast with other disordered materials, such as silica glasses or metallic glasses where a standard atomistic description is in principle possible, granular media are ruled by {\it mesoscale} frictional interactions between grains. As a consequence of this, frictional  granular materials can not be described by a Hamiltonian from which the inter-granule forces can be derived. The fundamental reason for this is that frictional forces depend on time and velocities, and can 
thus not be incorporated into a Hamiltonian. It was discovered and demonstrated recently that the
lack of a Hamiltonian has generic consequences for the dynamics of granular media in the form of oscillatory instabilities that can drive the system to catastrophic mechanical failure \cite{19CGPP,19CGPPa,19CCPP}. 

Since the dynamics of granular media is not derivable from a Hamiltonian, the usual approach to the stability of amorphous systems, which is based on the analysis of the Hessian matrix (second derivative of the Hamiltonian with respect to coordinates), is not tenable. Nevertheless, forces exist, and the dynamics follows Newton's equations for the accelerations in terms of these forces. The stability of a stationary solution of these equations is determined by the so-called ``J-matrix'' which is the first derivative of the forces with respect to the coordinates, cf. Sec.~\ref{model} below. The formalism that exposes the instability and its consequences were explored so far only in 2 dimensions \cite{19CGPP,19CGPPa}. In the present paper, we extend the formulation to 3 dimensions, compute analytically the J-matrix for assemblies of compressed frictional spheres subject to external shear forces, and demonstrate the instability and its consequences.

The structure of this paper is as follows: in Sect.~\ref{model} we describe the generalization of the model studied in Refs.~\cite{19CGPP,19CGPPa} to 3 dimensions. The force between spheres and the equations of motion are described. In Sect.~\ref{protocols} we discuss the numerical protocols used to expose the oscillatory instability. Section \ref{dynamics} describes the results of the numerical simulations and the catastrophic failure that results from the instability. The last section \ref{conclusions} offers a summary and some concluding remarks.

\section{Model and equations of motion}
\label{model}

\subsection{Forces}

The model discussed here consists of a binary mixture of $N$=100 frictional spheres of mass $m$ in a box of size $L^3$, half of which with radius $\sigma_1$=0.5 and the other half with $\sigma_2$=0.7. The position of the center of mass of the $i$th sphere is denoted $\B r_i$. The interaction between two spheres has a normal and a tangential component. When the assembly of spheres is compressed the spheres overlap. The normal force between the $i$th and the $j$th spheres is determined by the amount of overlap $\delta_{ij} \equiv \sigma_i+\sigma_j-r_{ij}$, where $\B r_{ij}\equiv \B r_i-\B r_j$. We choose for the normal force the Hertzian model, but stress that the qualitative nature of our results is independent of the precise choice of the forces:
\begin{equation}
\B F_{ij}^{(n)} = k_n \delta_{ij}^{3/2}\hat r_{ij} \ , \quad \hat r_{ij} \equiv \B r_{ij}/r_{ij}.
\label{Fn}
\end{equation}

The tangential force is caused by the tangential displacement $\B t_{ij}$ between adjacent spheres. The tangential force is always orthogonal to $\hat{r}_{ij}$. Upon first contact between the particles, $t_{ij}=0$. In three dimensions the tangential displacement is determined by a 3-dimensional angular coordinate ${\B \theta}_i\equiv \{\theta_i^x, \theta_i^y, \theta_i^z\}$. The change in tangential displacement is given by
\begin{equation}
d\B t_{ij} =d\B r_{ij} -(d\B r_{ij}\cdot \B r_{ij}) \hat r_{ij} +\hat r_{ij} \times (\sigma_i d\B\theta_i +\sigma_jd\B\theta_j) \ .
\end{equation}
Following this equation one computes $\B t_{ij}$ by integrating over time the relative velocity of the particles at the point of contact. In the Mindlin model, the tangential force depends on $\B t_{ij}$ and on  the contact area which is proportional to $\sqrt{\delta_{ij}}$~\cite{49Min}
\begin{equation}
\B F_{ij}^{(t)} = -k_t\delta_{ij}^{1/2}t_{ij} \hat t_{ij} \ .
\label{Min}
\end{equation}
Like in all frictional model the tangential force is required to satisfy the Coulomb condition
\begin{equation}
\B F_{ij}^{(t)} \le \mu \B F_{ij}^{(n)} \ ,
\label{Coul}
\end{equation}
where $\mu$ is the friction coefficient. To be able to compute the stability J-matrix we need to smooth out the Coulomb law such that the tangential force will have smooth derivatives; we choose:
\begin{eqnarray}
&&\B F_{ij}^{(t)} = -k_t\delta_{ij}^{1/2}\left[1+\frac{t_{ij}}{t^*_{ij}} -\left(\frac{t_{ij}}{t^*_{ij}}\right)^2\right]t_{ij} \hat t_{ij} \ , \nonumber\\
&&t^*_{ij} \equiv \mu \frac{k_n}{k_t} \delta_{ij} \ .
\label{Ft}
\end{eqnarray}
The derivative of the force with respect to $t_{ij}$ vanishes smoothly at $t_{ij}=t^*_{ij}$ and the Coulomb law Eq.~(\ref{Coul}) is fulfilled.

\subsection{Equations of motion}

Once we defined the forces we can write the equations of motion, which are simply Newton's equations
for an extended set of coordinates $\B q_i=\{\B r_i, \B \theta_i\}$:
\begin{eqnarray}
m\frac{d^2 \B r_{i}}{dt^2}&=&{\B F}_i(\B q_1,\B q_2,\cdots,\B q_N)\ , \nonumber\\
\label{Newton}
I_{i}\frac{d^2 \B \theta_{i}}{dt^2}&=&{\B T}_i(\B q_1,\B q_2,\cdots,\B q_N)\ ,
\end{eqnarray}
where  $I_i$ are moments of inertia for the spheres, ${\B F}_i$ are forces and ${\B T}_i$ are torques, respectively.

In the simulations reported below we employ a unit mass $m_i=1$ and moment of inertia $I_i = 0.4 m_i\sigma_i^2$.
The normal interaction between the granular particles is given by Eq.~(\ref{Fn}), while
the tangential one is given by Eq.~(\ref{Ft}), with $k_n=200000$ and $k_t=2k_n/7$.
We use $m$, $2\sigma_1$ and $\sqrt{m(2\sigma_1)^{-1/2}k_n^{-1}}$ as units of mass, length and time, respectively.
We fix the friction coefficient to a high value, $\mu = 10$, to emphasize that the existence of a Coulomb threshold is not responsible for the reported phenomenology.

\subsection{The stability matrix}

Using the smoothed out force Eq.~(\ref{Ft}) allows to define the stability matrix, which is an operator obtained from the derivatives of the force $\B F_i$ and the torque $\B T_i$ on each particle with respect to the coordinates. In other words
\begin{equation}
J_{ij}^{\alpha\xi} \equiv \frac{\partial \tilde F^\alpha_i}{\partial q_j^\xi}\ , \quad \tilde{ \B F}_i \equiv \sum_j \tilde{\B F}_{ij} \ ,
\end{equation}
where $\B q_j$ stands for either a spatial position or a tangential coordinate, and $\tilde {\B F}_i$ stands for either a force or a torque. We stress
the obvious fact that $\B J$ is not a symmetric operator.
Being real it can possess pairs of complex eigenvalues.
When these appear, the system will exhibit
oscillatory instabilities, since one of each complex pair will cause an oscillatory exponential divergence of any perturbation, and the other an oscillatory exponential decay.
The actual calculation of the operator $\B J$ for the 3-dimensional case is detailed in Appendix~\ref{Jacobian}.

\section{Simulation protocols and the birth of the instability}
\label{protocols}

The equations of motion are solved using two types of algorithms: ``Newtonian'' and ``Over-damped''. The first is simply a solution of the Newton equations of motion with the given forces, Eqs.~(\ref{Newton}).
The second algorithm is solving the same equations of motion but with a damping force that is proportional to the velocities of the centers of mass $\dot{\B r_i}$ of the spheres with a coefficient of proportionality $\eta_v=m\eta_0$.
If not otherwise mentioned we use $\eta_0 = 2.2\times 10^{-2}$ expressed in reduced units.
This value of $\eta_0$ ensures that the dynamics is over-damped as the damping timescale $\eta_v^{-1}$ is of the order of the time that sounds needs to travel one particle diameter.
The reader should note that the over-damped equations are employed to reach a stationary solutions with zero forces $\B F_i$ and torques $\B T_i$ on all the disks for the purpose of computing the J-matrix.
We use LAMMPS~\cite{Plimpton1995} to perform the numerical integration for these two algorithms,
with integration timestep $10^{-5}\sqrt{(2\sigma_1)^{1/2}k_nm^{-1}}$.
\begin{figure}
	\includegraphics[width=0.8\columnwidth] {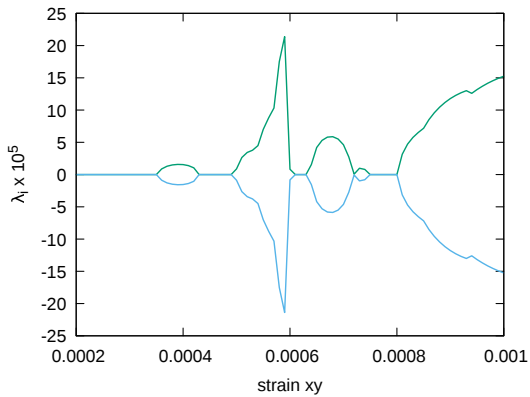}
	\caption{Evolution of minimum (blue line) and maximum (green line) complex eigenvalue pair with increasing strain, using AQS protocol. }
	\label{bifurcation}
\end{figure}

 An initial configuration is prepared by arranging assemblies of binary spheres randomly in a three dimensional box and then perform two consecutive runs of overdamped dynamics to bring the configuration at mechanical equilibrium.
The initial configuration is prepared focusing on a frictionless system (i.e. $\mu=0$), and hence has no complex eigenvalues.
Afterwards, we switch on friction, and perform athermal quasi static (AQS) simulations: starting from the initial stable configuration we shear the simulation box along the  ($x$) direction by the amount $\delta\gamma$ and then we run the overdamped dynamics until the system reaches mechanical equilibrium. The system is considered to be in mechanical equilibrium when the net force on each sphere is less than $5 \cdot 10^{-14}$. After each such steps we diagonalize the J-matrix and calculate the eigenvalues.
As in the 2D system \cite{19CGPP}, at some value of the strain we identify the birth of a couple of conjugate complex eigenvalues.
\begin{figure}
\vskip 0.5 cm
\includegraphics[width=0.8\columnwidth] {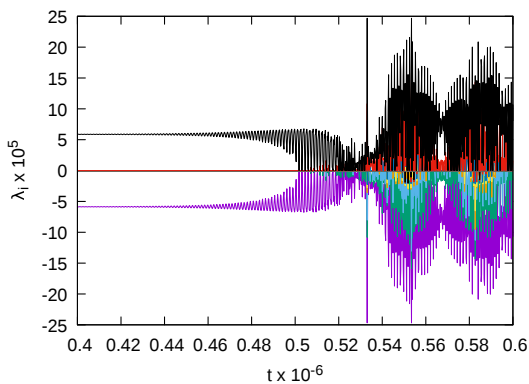}
\caption{Time dependence of the imaginary component of all the 600 eigenvalues of the system, during a Newtonian simulation.}
\label{evolution}
\end{figure}

\subsection{The oscillatory instability}
When a pair of complex eigenvalues $\lambda_{1,2}=\lambda_r\pm i\lambda_i$ gets born, a novel instability mechanism develops.
A pair of complex conjugate eigenvalues correspond to FOUR solutions $e^{i\omega t}$ to the linearized equation
of motion with
\begin{equation}
i\omega_{1,2} = \omega_i \pm i\omega_r\ , \quad i\omega_{3,4}=-\omega_i \pm i\omega_r\ ,
\label{four}
\end{equation}
with $\omega_r \pm i\omega_i = \sqrt{\lambda_r \pm i\lambda_i}$.
The first pair in Eq.~(\ref{four}) will induce an oscillatory motion with an exponential growth of any deviation $\B q(0)$ from a state of mechanical equilibrium,
\begin{equation}
\B q(t) = \B q(0) e^{\omega_i t} \sin(\omega_r t).
\label{growth}
\end{equation}
The second pair represents an exponentially decaying oscillatory solution. The actual spatial dynamics that sets in due to this instability will be discussed below in Sect.~\ref{dynamics}.
Figure~\ref{bifurcation} shows the imaginary component of the dominant eigenpair. Note that with increasing strain the imaginary component can
reduce and disappear, then reappearing as $\gamma$ is increased further. At higher values of $\gamma$ one can easily obtain the simultaneous existence of many complex eigenpairs.
\begin{figure}
	\includegraphics[width=0.9\columnwidth] {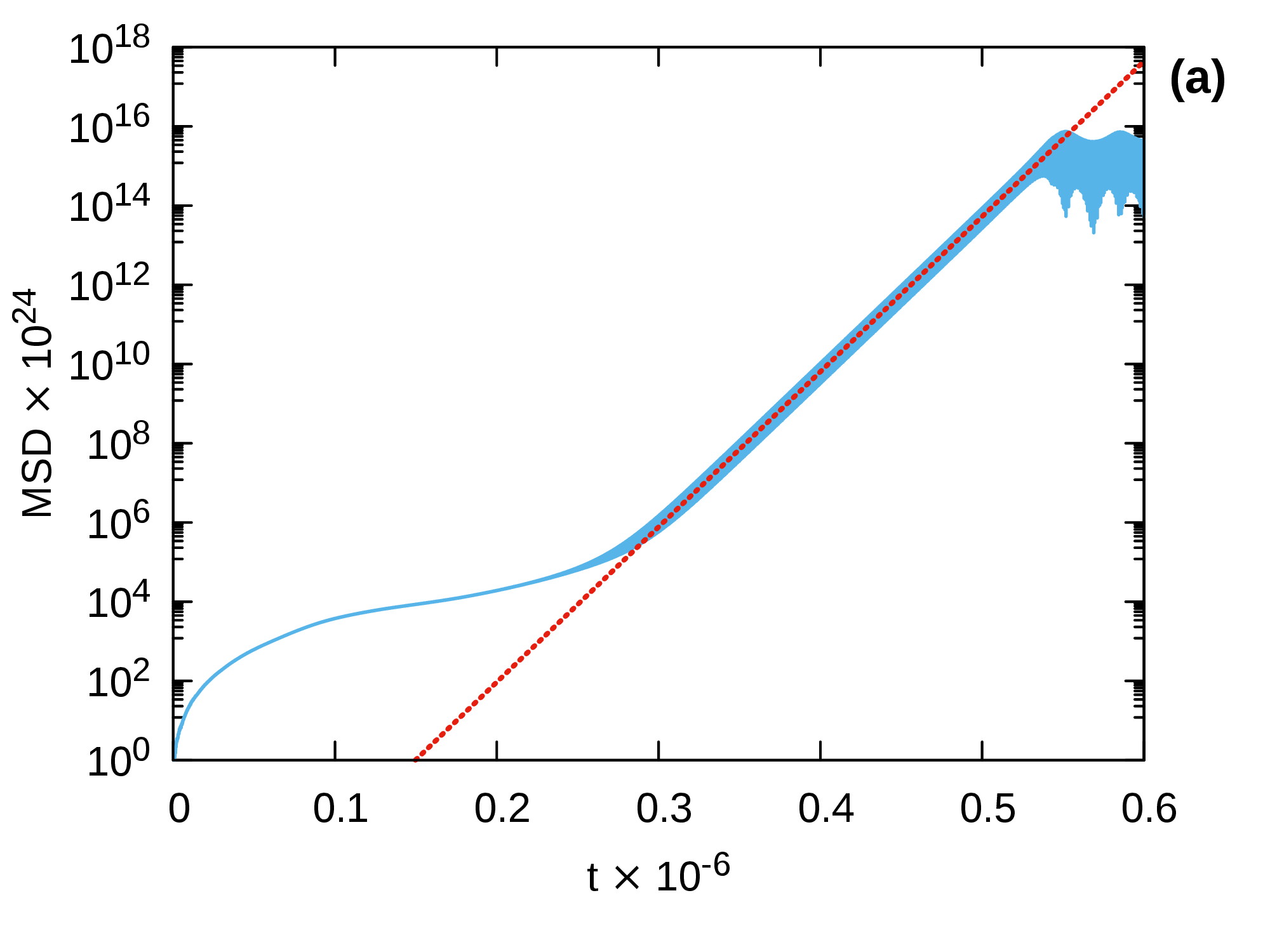}
	\includegraphics[width=0.92\columnwidth] {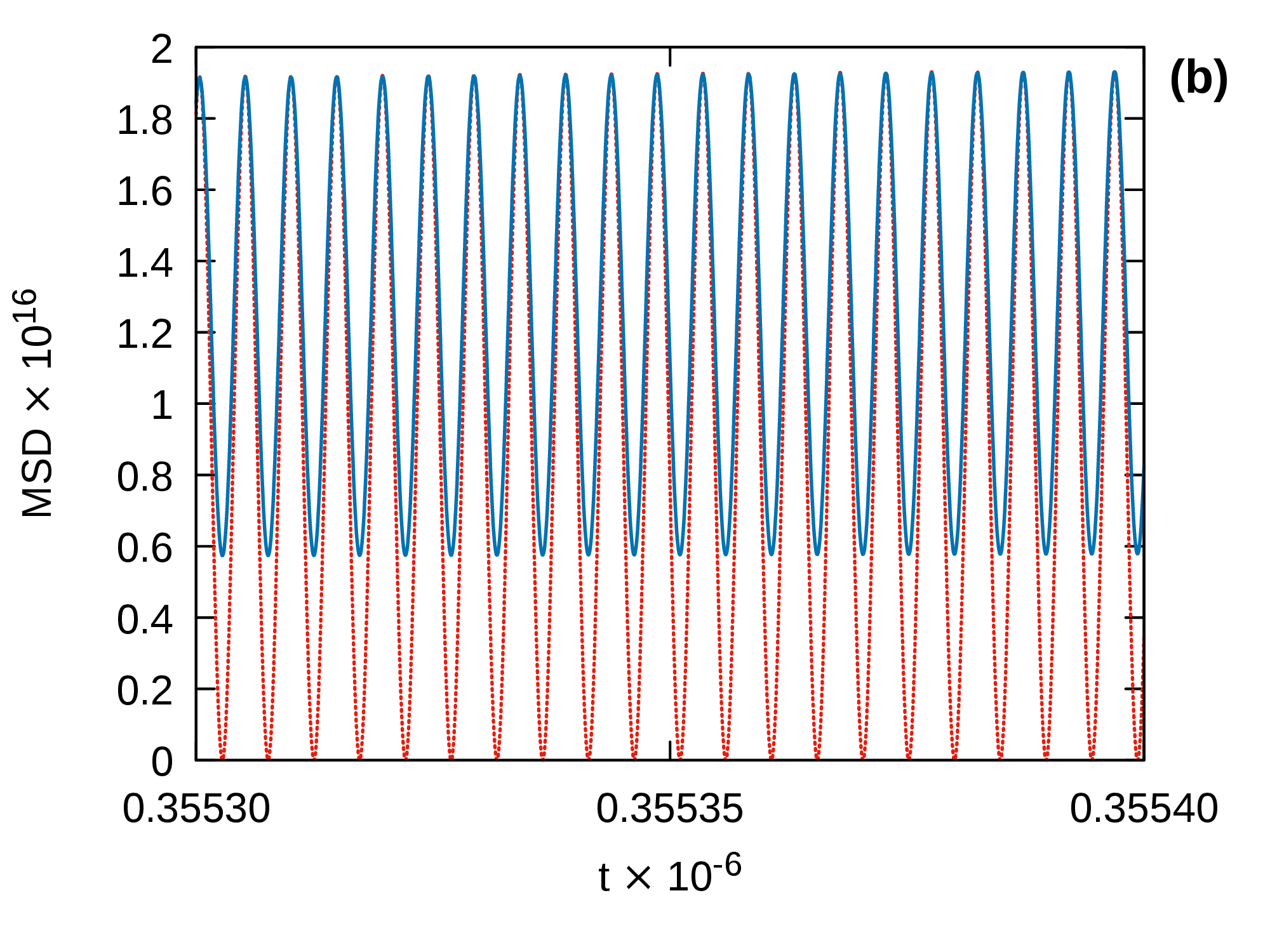}
	\caption{MSD (blue line) and fit (red dotted line) during the Newtonian dynamics. In (a) the fit is the predicted exponential growth from the linear instability, $a_0 \exp[2\omega_i t]$, with $a_0$ being fitted. In (b) the MSD is fitted by the exponential oscillatory instability prediction, $a_0 \exp[2\omega_i t][\sin(\omega_r t + \psi)]^2$ with $\psi$ fitted, as explained in the main text.}
	\label{MSD}
\end{figure}

\section{Dynamical Consequences of the Instability}
\label{dynamics}

Once the J-matrix exhibits at least one conjugate complex pair of eigenvalues, the system loses mechanical stability. To see the evolution
under the influence of this instability one needs to run the Newtonian equations Eqs.~(\ref{Newton}).
Starting from a configuration with only a pair of complex eigenvalues, with large $\lambda_i=5.87\times10^{-5}$, we run the Newtonian dynamics and evaluate the eigenvalues of the J-matrix at fixed intervals of time. As in the 2D case, we observe that the eigenvalues remain constant for a period of time until an instability develops, cf. (Figure~\ref{evolution}).
The complex eigenvalue induce a spiral motion in 3D as shown below.

To underline the exponential growth of small perturbations we consider the mean-square displacement $M(t)$ as a function of time:
\begin{eqnarray}
M(t) & \equiv& \frac{1}{N} \sum_i \big[\Delta r_i^x (t)^2 + \Delta r_i^y (t)^2 + \Delta r_i^z (t)^2 \nonumber \\
&+& \sigma_i^2 \big( \Delta \theta^x_i (t)^2+ \Delta \theta^y_i (t)^2 + \Delta \theta^z_i (t)^2 \big)\big]  \,
\end{eqnarray}
which is reported in Figure~\ref{MSD} (a).
We observe an increase in time of about 10 orders of magnitude following an oscillatory exponential growth. The blue curve represents the computed MSD as a function of time, the red dotted curve is the predicted exponential: $a_0 \exp[2\omega_i t]$ where $a_0$ represent an offset constant. Panel~(b) reports a blow up of the MSD growth with the fitted function (red dotted curve) being $a_0 \exp[2\omega_i t][\sin(\omega_r t + \psi)]^2$ with $\psi$ fitted. The values of fitted $\omega_r$ and $\omega_i$ correspond perfectly to those expected
frequencies computed from the complex eigenvalue.

\begin{figure}
  \includegraphics[scale=0.37] {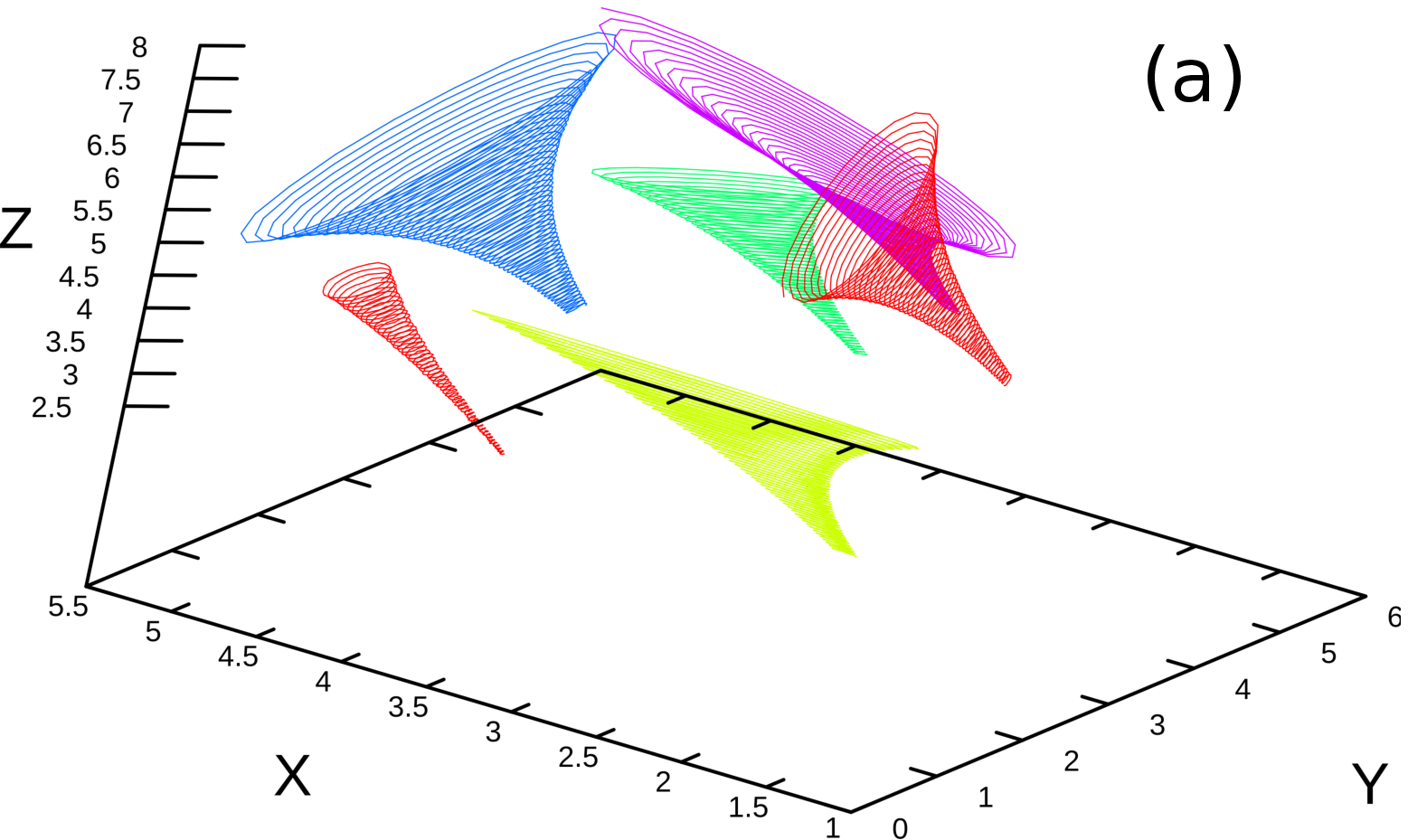}
  \includegraphics[scale=0.37] {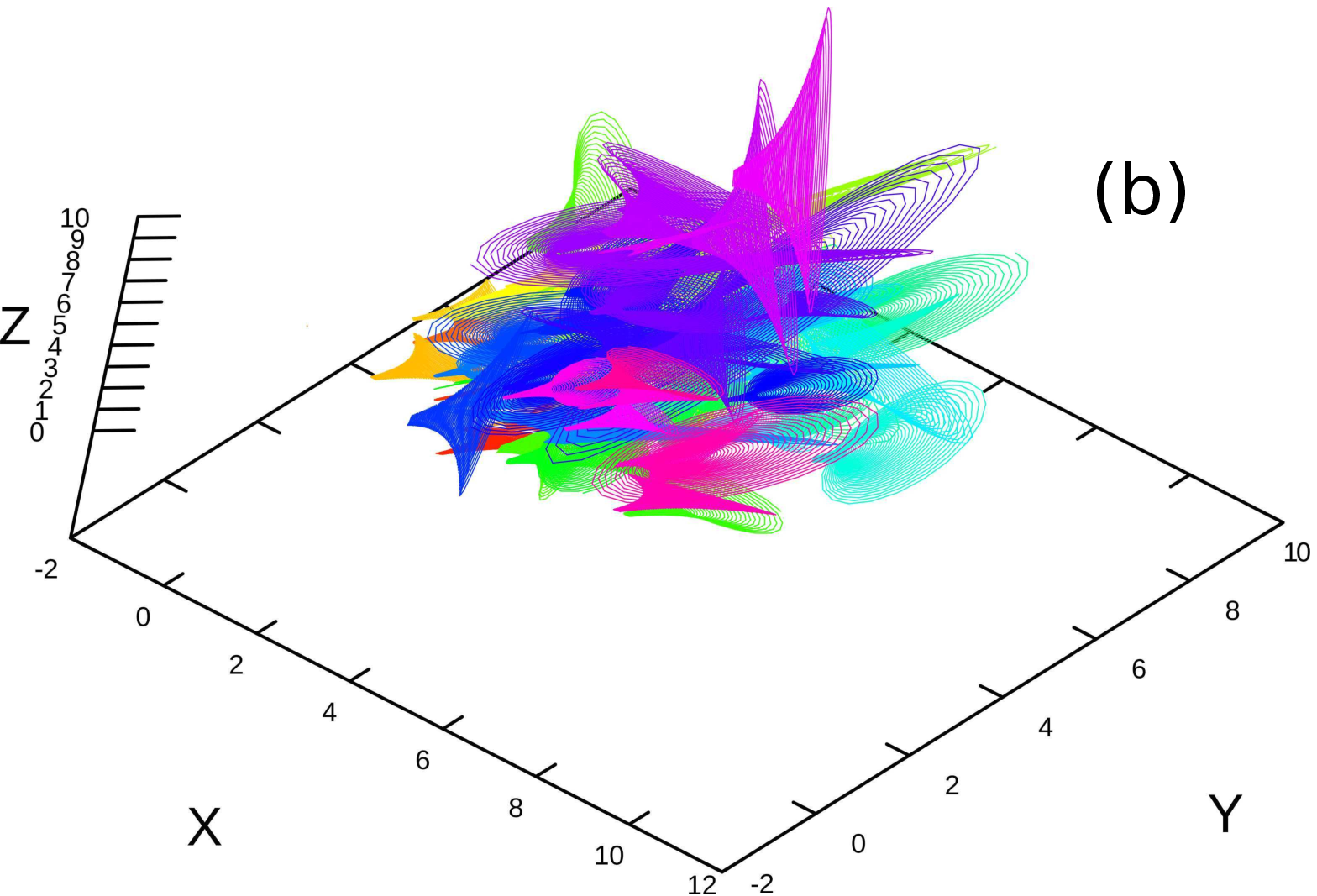}
\caption{Views of the trajectory of (a)~few and (b)~many spheres during the Newtonian dynamics. Here actual particle displacements are amplified by a factor $10^9$. The view from a specific perspective might resemble the spirals in 2D.}
\label{spiral}
\end{figure}

The investigation of a particle trajectory during the development of the instability has a spiral motion, see Figure~\ref{spiral}.

\begin{figure}
\includegraphics[width=0.7\columnwidth] {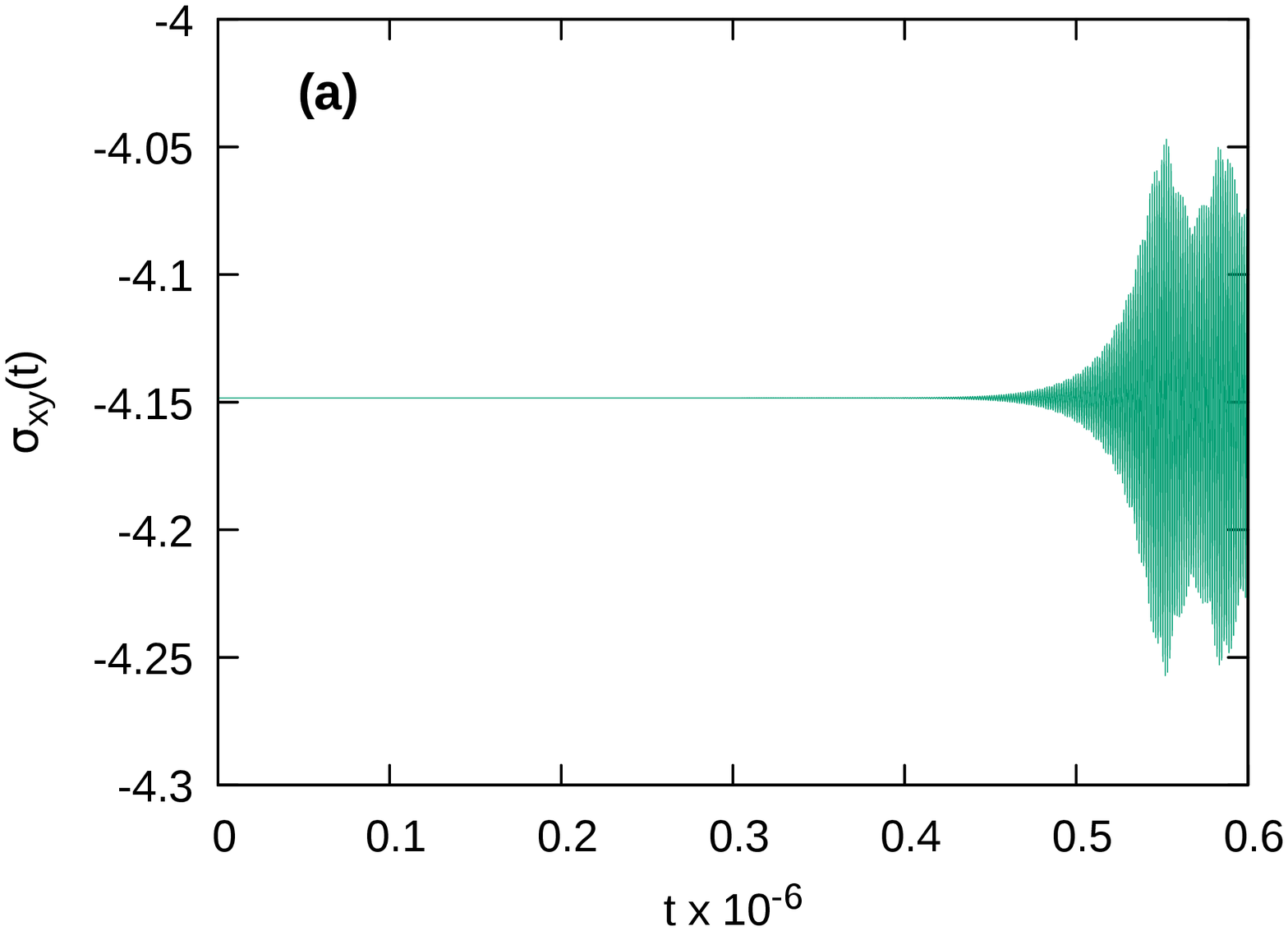}
\vspace{3mm}
\includegraphics[width=0.67\columnwidth] {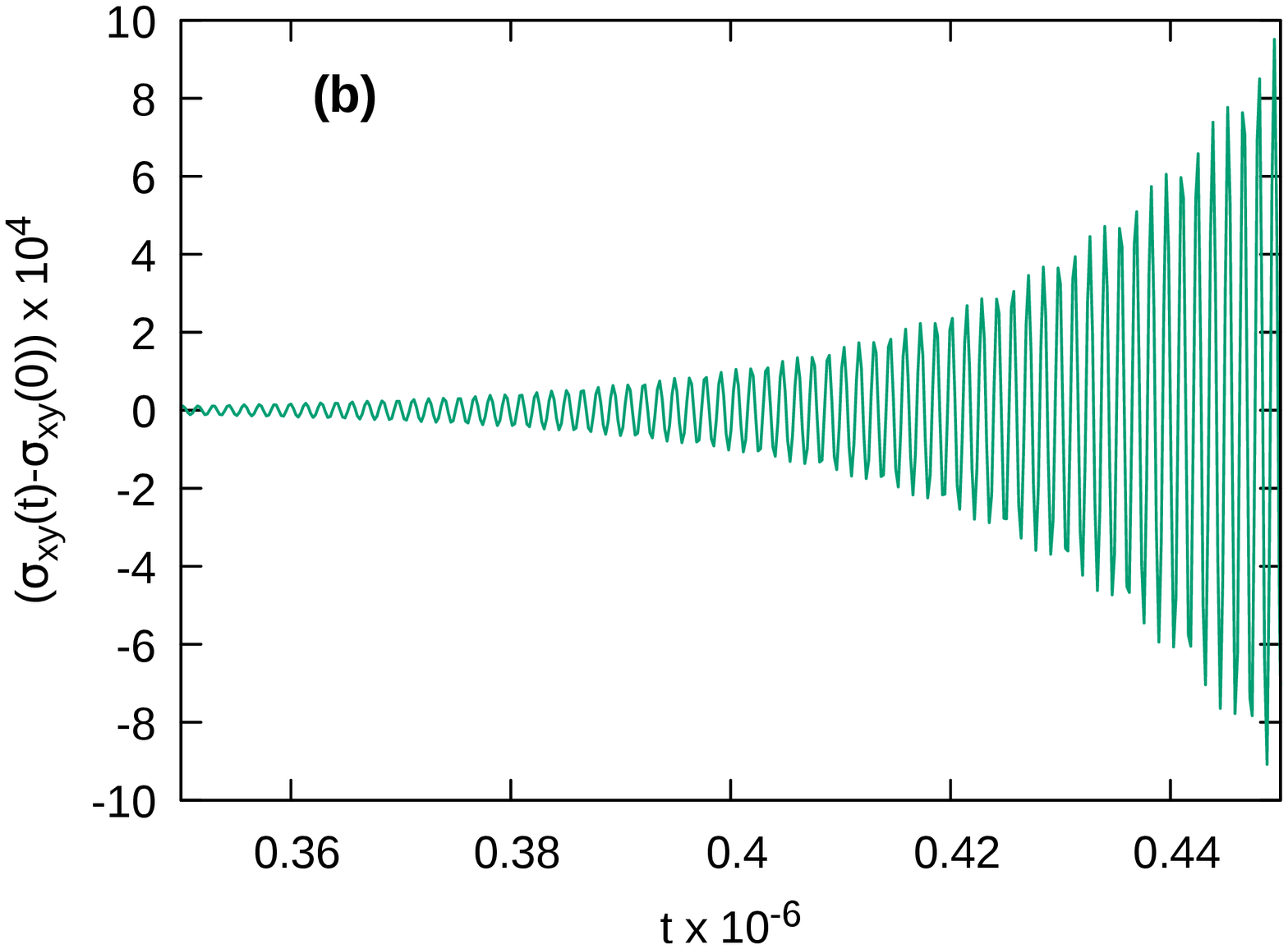}
\caption{(a) The evolution of the shear stress (virial contribution) $\sigma_{xy}$ versus time, during Newtonian dynamics. (b)~Enlarged view of the stress change during the instability.} 
\label{stress}
\end{figure}

We focus finally on the virial component of the shear stress $\sigma_{xy}=-\frac{1}{L^2} \sum_{i\neq j} r_{ij}^x F_{ij}^y$ and find that the trend reported in Figure~\ref{stress} is also similar to the 2D case.
\vskip 1.0 cm

\section{Conclusions}
\label{conclusions}
We presented the development of mechanical instabilities in a disordered packing of frictional spheres, extending to three dimensions the description of the J-matrix previously derived for two-dimensional disks \cite{19CGPP,19CGPPa}. We have shown that there exist instabilities arising in typical granular compounds under external shear which are formally related to the emergence of imaginary eigenvalues; these dictate the characteristic time of the exponential growth and of the oscillatory period of particles motion. Importantly, these results should hold for any kind of normal and tangential force expressions which are not derivable from a Hamiltonian, thus providing a general framework to predict the shear force limit and the response of the system upon its crossing.

The mechanism we discuss here should describe the amplification of mechanical perturbations in countless granular matter systems extending over a large range of scales.  Earthquake faults are often separated by a granular grit that is sheared by the slow motion of tectonic plates \cite{johnson2005nonlinear}. The development and growth of mechanical instabilities described here has thus straightforward application to slip nucleation in the geophysical context \cite{marone1998laboratory,johnson2013acoustic}. Other relevant application of our theoretical framework comprise the design of pharmaceutical tablets, the modeling of the formation of icebergs, or the stability of grains in silos. As long
as the dynamics is describable by forces which are not derivable from a Hamiltonian, this instability should be generic. 

\section*{Acknowledgements}{This work is supported by the scientific and cooperation agreement between Italy and Israel through the project COMPAMP/DISORDER, by the ISF-Singapore exchange program and by the US-Israel Binational Science Foundation.We thank Massimo Pica Ciamarra for his important contributions to the development of the ideas expounded in this paper.}
\appendix
\begin{widetext}

\section{Calculation of the operator $\mathbf{\textit{J}}$ in 3D~\label{Jacobian}}
The Jacobian operator ${\B J}$, which represents the dynamical response of the system, is given by the derivative of the forces and of the torques acting on the particles with respect to all the degrees of freedom. The interaction forces used in this work is recalled in Sec.~\ref{sec:forces}, the tangential displacement and its derivative are described in Sec.~\ref{sec:tangential}. 
The expressions for all the components of ${\B J}$ and how these components are arranged as a matrix
are reported in Sec.~\ref{sec:J}.

\subsection{Interaction force~\label{sec:forces}}
In our model, a pair of granules interacts when they overlap. The overlap distance $\delta_{ij}$ is given by
\begin{equation}
  \delta_{ij}=\sigma_i+\sigma_j-r_{ij},
  \label{Eq:delta}
\end{equation}
where $r_{ij}$ is the center-to-center distance of a pair-$i$ and $j$, and $\sigma_i$ is the radius of particle-$i$.
The pair vector ${\B r}_{ij}$ is defined as
\begin{equation}
  {\B r}_{ij}=\B r_i - \B r_j.
  \label{Eq:rij}
\end{equation}
The pair-interaction force ${\B F}_{ij}$ has two contributions. ${\B F}^{(n)}_{ij}$ is the force acting along the normal direction of the pair $\hat{r}_{ij}$, and ${\B F}^{(t)}_{ij}$ is the force acting along the tangential direction of the pair ${\hat t}_{ij}$.
The normal force is Hertzian:
\begin{equation}
  {\B F}^{(n)}_{ij}= k_n\delta^{3/2}_{ij} {\hat r}_{ij},
  \label{Eq:Fn}
\end{equation}
where $k_n$ is the force constant with dimension: Force per length${}^{3/2}$.
The tangential force ${\B F}^{(t)}_{ij}$ is a function of both the overlap distance $\delta_{ij}$ and the tangential displacement ${\B t}_{ij}$.
As done in the previous work for 2D frictional system, we have modified the standard expression for ${\B F}^{(t)}_{ij}$ and included a few higher order terms of $t_{ij}$ (i.e., $|{\B t}_{ij}|$) such that the derivative of the force function $F^{(t)}_{ij}$ with respect to tangential distance $t_{ij}$ becomes continuous and it goes to zero smoothly.
We use the following form:
\begin{equation}
  \label{Eq:Fs}
  \begin{split}
    {\B F}^{(t)}_{ij} & = -k_t\delta^{1/2}_{ij}\left[1 + \frac{t_{ij}}{t^*_{ij}} - \left(\frac{t_{ij}}{t^*_{ij}}\right)^2 \right] t_{ij}{\hat t}_{ij}\\
    & = -k_t \delta^{1/2}_{ij} t^*_{ij}{\hat t}_{ij}, \ \ \ \textrm{if}\ \ k_t\delta^{1/2}_{ij}t_{ij} > \mu |{\B F}^{(n)}_{ij}|,
  \end{split}
\end{equation}
where $k_t$ is the tangential force constant. Its dimension is force per length${}^{3/2}$. $t^*_{ij}$ is the threshold tangential distance:
\begin{equation}
t^*_{ij} = \mu \frac{k_n}{k_t} \delta_{ij},
\label{Eq:s*}
\end{equation}
where $\mu$ is the friction coefficient, a scalar quantity, which essentially determines the maximum strength of the tangential force with respect to the normal force at a fixed overlap $\delta_{ij}$. The derivative of $F^{(t)}_{ij}$ with respect to $t_{ij}$ vanishes at $t^*_{ij}$, as it turns out
\begin{equation}
 \label{Eq:dFsds}
  \begin{split}
    \frac{\partial F^{(t)}_{ij}}{\partial t_{ij}} & = k_t\delta^{1/2}_{ij}\left[1 + 2\frac{t_{ij}}{t^*_{ij}} - 3\left(\frac{t_{ij}}{t^*_{ij}}\right)^2 \right] \\
    & = 0, \ \ \ \textrm{if}\ \ k_t \delta^{1/2}_{ij}t_{ij} > \mu |{\B F}^{(n)}_{ij}|.
  \end{split}
\end{equation}

{\bf We stress here that the above forces imply a non Hamiltonian dynamics.} That is, there is not
a function $U(\delta,t)$ such that $F^{(n)} = -\frac{\partial U}{\partial \delta}$ and
$F^{(t)} = -\frac{\partial U}{\partial t}$.

\subsection{Tangential displacement\label{sec:tangential}}
The tangential force is a function of both $\B t_{ij}$ and ${\B r_{ij}}$. The derivative of this force thus includes the derivative of the two latter quantities. Here we evaluate these derivatives
using the chain rule.

The derivative of tangential displacement ${\B t}_{ij}$ with respect to time $t$ is
\begin{equation}
  \frac{\mathrm d {\B t}_{ij}}{\mathrm d t} = {\B v}_{ij} - {\B v}^n_{ij} + \hat{r}_{ij}\times(\sigma_i{\B \omega}_i + \sigma_j{\B \omega}_j),
  \label{Eq:dsdt}
\end{equation}
where ${\B v}_{ij}={\B v}_i-{\B v}_j$ is the relative velocity of pair-$i$ and $j$. ${\B v}^n_{ij}$ is the projection of ${\B v}_{ij}$ along the normal direction $\hat{r}_{ij}$. ${\B v}_{ij} - {\B v}^n_{ij}$ is the tangential component of the relative velocity.
${\B \omega}_i$ and ${\B \omega}_j$ are the angular velocity of $i$ and $j$, respectively.
In differential form, the above equation reads:
\begin{equation}
  \mathrm d {\B t}_{ij} =  \mathrm d{\B r}_{ij} -  (\mathrm d{\B r}_{ij}\cdot\hat{r}_{ij})\hat{r}_{ij} + \hat{r}_{ij}\times(\sigma_i\mathrm d{\B \theta}_i + \sigma_j\mathrm d{\B \theta}_j),
  \label{Eq:ds}
\end{equation}
where $\mathrm d{\B \theta}_i$ is the angular displacement of $i$ which follows the relation: $\mathrm d{\B \omega}_i = \frac{\mathrm d {\B \theta}_i}{\mathrm d t}$.

Here on, we assume the three-dimensional ({$\bf 3D$}) system.
Therefore, $\omega_i$, and so $\theta_i$, have components along $\hat{x},\hat{y},\hat{z}$, and
the cross product is
\begin{equation}
\hat{r}_{ij}\times \mathrm{d}\boldsymbol{\theta}_i = 
\left(\frac{y_{ij}}{r_{ij}}\mathrm{d}\theta_i^z - \frac{z_{ij}}{r_{ij}}\mathrm{d}\theta_i^y\right)\hat{x}
 + \left(\frac{z_{ij}}{r_{ij}}\mathrm{d}\theta_i^x - \frac{x_{ij}}{r_{ij}}\mathrm{d}\theta_i^z \right)\hat{y}
 + \left(\frac{x_{ij}}{r_{ij}}\mathrm{d}\theta_i^y - \frac{y_{ij}}{r_{ij}}\mathrm{d}\theta_i^x\right)\hat{z}
\label{Eq:crossprod}
\end{equation}
Now if particle-$i$ changes its position the angular displacement remains unaffected, i.e.~$\frac{\mathrm d \theta_i}{\mathrm d r^\alpha_i}=0$. Thus, the change in tangential displacement along $\beta$ due to the change in position of particle-$i$ along $\alpha$ only contributes in translations, and it can be written as
\begin{equation}
  \frac{\mathrm d t^\beta_{ij}}{\mathrm d r^\alpha_i} = \Delta_{\alpha\beta} - \frac{r^\alpha_{ij} r^\beta_{ij}}{{r}^2_{ij}},
  \label{Eq:dsbdra}
\end{equation}
where $\Delta_{\alpha\beta}$ is the Kronecker delta which is one when $\alpha=\beta$, or else zero. Similarly, a change in rotational coordinates does not modify the particles relative distance, i.e.~$\frac{\mathrm d r^\beta_{ij}}{\mathrm d \theta_i}=0$. Thus, the change in tangential displacement along $\beta$ due to the change in $\theta_i$ is
\begin{equation}
\begin{bmatrix}
\frac{\dd t_{ij}^x}{\dd \theta_i^x}=0                               ~~&~~ \frac{\dd t_{ij}^x}{\dd \theta_i^y}=-\sigma_i \frac{z_{ij}}{r_{ij}} ~~&~~ \frac{\dd t_{ij}^x}{\dd \theta_i^z}=+\sigma_i \frac{y_{ij}}{r_{ij}} \\
\frac{\dd t_{ij}^y}{\dd \theta_i^x}=+\sigma_i\frac{z_{ij}}{r_{ij}}  ~~&~~ \frac{\dd t_{ij}^y}{\dd \theta_i^y}=0                               ~~&~~ \frac{\dd t_{ij}^y}{\dd \theta_i^z}=-\sigma_i \frac{x_{ij}}{r_{ij}} \\
\frac{\dd t_{ij}^z}{\dd \theta_i^x}=-\sigma_i \frac{y_{ij}}{r_{ij}} ~~&~~ \frac{\dd t_{ij}^z}{\dd \theta_i^y}=+\sigma_i \frac{x_{ij}}{r_{ij}} ~~&~~ \frac{\dd t_{ij}^z}{\dd \theta_i^z}=0 \\
\end{bmatrix}
 \label{Eq:dsbdtheta}
\end{equation}
Now the magnitude of tangential distance $t_{ij}$ can be obtained from the relation $t^2_{ij} = \sum_{\alpha} {t^\alpha_{ij}}^2$. Its differential follows $\mathrm d t_{ij} =  \sum_{\alpha} \frac{t^\alpha_{ij}}{t_{ij}} \mathrm d t^\alpha_{ij}$. The derivatives of tangential distance $t_{ij}$ with respect to $r^\alpha_{i}$ and $\theta^\alpha_i$ can be expressed as
\begin{eqnarray}
  \label{Eq:dsmagdra}
  \frac{\mathrm d t_{ij}}{\mathrm d r^\alpha_i} &=&  \left(\frac{t^x_{ij}}{t_{ij}}\right)\frac{\mathrm d t^x_{ij}}{\mathrm d r^\alpha_i} + \left(\frac{t^y_{ij}}{t_{ij}}\right)\frac{\mathrm d t^y_{ij}}{\mathrm d r^\alpha_i}+ \left(\frac{t^z_{ij}}{t_{ij}}\right)\frac{\mathrm d t^z_{ij}}{\mathrm d r^\alpha_i}, \\
  \frac{\mathrm d t_{ij}}{\mathrm d \theta^\alpha_i} &=&  \left(\frac{t^x_{ij}}{t_{ij}}\right)\frac{\mathrm d t^x_{ij}}{\mathrm d \theta^\alpha_i} + \left(\frac{t^y_{ij}}{t_{ij}}\right)\frac{\mathrm d t^y_{ij}}{\mathrm d \theta^\alpha_i}+ \left(\frac{t^z_{ij}}{t_{ij}}\right)\frac{\mathrm d t^z_{ij}}{\mathrm d \theta^\alpha_i}.
  \label{Eq:dsmagdtheta}
\end{eqnarray}
With the help of equations~(\ref{Eq:dsbdra}) and~(\ref{Eq:dsbdtheta}) we can solve the above two differential equations. As the tangential threshold is a linear function of overlap distance $\delta_{ij}$ (see~(\ref{Eq:s*})), it also gets modified due to a change in $r^\alpha_i$ as
\begin{equation}
  \frac{\mathrm d t^*_{ij}}{\mathrm d r^\alpha_i} = - \mu\left(\frac{k_n}{k_t}\right)\frac{r^\alpha_{ij}}{r_{ij}},
  \label{Eq:ds*dra}
\end{equation}
and it is unaffected by the change in rotation, i.e.~$\frac{\mathrm d t^*_{ij}}{\mathrm d \theta^\alpha_i}=0$.

\section{Evaluation of $\mathbf{\textit{J}}$ \label{sec:J}}
\subsection{Derivative of tangential force}
The derivative of tangential force (equation~(\ref{Eq:Fs})) with respect to $r^\alpha_{i}$:
\begin{eqnarray}
  \frac{\partial {F^{(t)}_{ij}}^\beta}{\partial r^\alpha_i} &=& -k_t\frac{\partial}{\partial r^\alpha_i}\left[\delta^{1/2}_{ij}\left(t^\beta_{ij} + \tilde{t} t^\beta_{ij} - {\tilde t}^2t^\beta_{ij} \right)\right]
  \nonumber \\
  &=& -\frac{1}{2}\delta^{-1}_{ij}\frac{r^\alpha_{ij}}{r_{ij}}{F^{(t)}_{ij}}^\beta - k_t\delta^{1/2}_{ij} \left[ (1+{\tilde t}-{\tilde t}^2)\frac{\partial t^\beta_{ij}}{\partial r^\alpha_i} + ({\tilde t}^\beta - 2{\tilde t}{\tilde t}^\beta)\frac{\partial t_{ij}}{\partial r^\alpha_i} + (-{\tilde t}{\tilde t}^\beta+2{\tilde t}^2{\tilde t}^\beta)\frac{\partial t^*_{ij}}{\partial r^\alpha_i} \right]
  \label{Eq:dFsbdra}
\end{eqnarray}
Here we use the notation $\tilde t$ to represent the ratio $t_{ij}/t^*_{ij}$, and the notation ${\tilde t}^\beta$ for ${t_{ij}}^\beta/t^*_{ij}$. The expressions for all the three partial differentiation in~(\ref{Eq:dFsbdra}) are already shown in~(\ref{Eq:dsbdtheta}), (\ref{Eq:dsmagdra}), and (\ref{Eq:ds*dra}).

Similarly, the derivative of tangential force with respect to $\theta_{i}^\alpha$ (using the same notation as above) can be found as
\begin{equation}
    \frac{\partial F_{ij}^{(t)}{}^\beta}{\partial \theta_i^\alpha}=-k_t \delta_{ij}^{1/2} \left[ (1+ \tilde{t}-\tilde{t}^2) \frac{\partial t_{ij}^\beta}{\partial \theta_i^\alpha} + (\tilde{t}^\beta-2\tilde{t}\tilde{t}^\beta) \frac{\partial t_{ij}}{\partial \theta_i^\alpha} \right]
    \label{Eq:dFsbdtheta}
\end{equation}
From the above two equations it is then understood that if ${\B r}_{ij}$ and ${\B t}_{ij}$ are known the differential equations can be solved easily. When ${\tilde t}^\beta$ is negligible for all $\beta$, then ${\tilde t}\approx 0$. 
This translates to 
$-k_t \delta_{ij}^{1/2} \frac{\partial t_{ij}^\beta}{\partial \theta_i^\alpha}$
implying that even in the case of zero tangential displacement and therefore, zero tangential force, the above derivative can be finite.

\subsection{Derivative of normal force}
The derivative of normal force (equation~(\ref{Eq:Fn})) with respect to $r^\alpha_{i}$:
\begin{eqnarray}
  \frac{\partial {F^{(n)}_{ij}}^\beta}{\partial r^\alpha_i} &=& k_n \frac{\partial }{\partial r^\alpha_i} \left[ \delta^{3/2}_{ij} \frac{r^\beta_{ij}}{r_{ij}} \right]
  \nonumber \\
  &=& k_n \delta^{1/2}_{ij}\left[ \Delta_{\alpha\beta}\frac{\delta_{ij}}{r_{ij}} - \frac{3}{2}\frac{r^\alpha_{ij} r^\beta_{ij}}{r^2_{ij}} -\left(\frac{\delta_{ij}}{r_{ij}}\right) \frac{r^\alpha_{ij} r^\beta_{ij}}{r^2_{ij}} \right],
  \label{Eq:dFnbdra}
\end{eqnarray}
where $\Delta_{\alpha\beta}$ is the Kronecker delta. The derivative of total force which reads:
\begin{eqnarray}
\label{Eq:dFbdra}
\frac{\partial {F_{ij}}^\beta}{\partial r^\alpha_i} &=& \frac{\partial {F^{(n)}_{ij}}^\beta}{\partial r^\alpha_i} + \frac{\partial {F^{(t)}_{ij}}^\beta}{\partial r^\alpha_i} \\
\frac{\partial {F_{ij}}^\beta}{\partial \theta^\alpha_i}  &=& \frac{\partial {F^{(t)}_{ij}}^\beta}{\partial \theta^\alpha_i}
\label{Eq:dFbdtheta}
\end{eqnarray}
can be solved using~(\ref{Eq:dFnbdra}),~(\ref{Eq:dFsbdra}), and~(\ref{Eq:dFsbdtheta}).

\subsection{Derivative of Torque}
The torque of particle-$j$ due to tangential force ${\B F^{(t)}}_{ij}$ is ${\B T}_j = -\sigma_j\left({\hat r}_{ij} \times {\B F^{(t)}}_{ij}\right) \equiv \sigma_j \tilde{\B{T}}_{ij}$.
In 3D, the components of $\tilde{\B T}_{ij}$ are:
\begin{equation}
\begin{split}
\widetilde{T}_{ij}^{x}&=-\left[ \left( \frac{y_{ij}}{r_{ij}}\right) F^{(t)}_{ij}{}^{z} - \left(  \frac{z_{ij}}{r_{ij}} \right) F^{(t)}_{ij}{}^{y}  \right] \\
\widetilde{T}_{ij}^{y}&=-\left[ \left( \frac{z_{ij}}{r_{ij}}\right) F^{(t)}_{ij}{}^{x} - \left(  \frac{x_{ij}}{r_{ij}} \right) F^{(t)}_{ij}{}^{z}  \right]  \\
  \widetilde{T}_{ij}^{z}&=-\left[ \left( \frac{x_{ij}}{r_{ij}}\right) F^{(t)}_{ij}{}^{y} - \left(  \frac{y_{ij}}{r_{ij}}\right) F^{(t)}_{ij}{}^{x} \right].
  \label{Eq:Tz}
\end{split} 
\end{equation}
The derivative of ${\tilde T}_{ij}^x,~{\tilde T}_{ij}^y,~{\tilde T}_{ij}^z$ then are:
\begin{equation}
  \label{Eq:dTdra}
  \begin{split}
\frac{\partial \widetilde{T}_{ij}^{x}}{\partial r_i^\alpha} & = - \left( \frac{\delta_{\alpha y}}{r_{ij}} - \frac{y_{ij}r_{ij}^\alpha}{r_{ij}^3} \right) F^{(t)}_{ij}{}^z - \left( \frac{y_{ij}}{r_{ij}} \right) \frac{\partial F^{(t)}_{ij}{}^z}{\partial r^{\alpha}_{i}} +
\left( \frac{\delta_{\alpha z}}{r_{ij}} - \frac{z_{ij}r_{ij}^\alpha}{r_{ij}^3} \right) F^{(t)}_{ij}{}^y + \left( \frac{z_{ij}}{r_{ij}} \right) \frac{\partial F^{(t)}_{ij}{}^y}{\partial r^{\alpha}_{i}} \\
\frac{\partial \widetilde{T}_{ij}^{y}}{\partial r_i^\alpha} & = - \left( \frac{\delta_{\alpha z}}{r_{ij}} - \frac{z_{ij}r_{ij}^\alpha}{r_{ij}^3} \right) F^{(t)}_{ij}{}^x - \left( \frac{z_{ij}}{r_{ij}} \right) \frac{\partial F^{(t)}_{ij}{}^x}{\partial r^{\alpha}_{i}} +
\left( \frac{\delta_{\alpha x}}{r_{ij}} - \frac{x_{ij}r_{ij}^\alpha}{r_{ij}^3} \right) F^{(t)}_{ij}{}^z + \left( \frac{x_{ij}}{r_{ij}} \right) \frac{\partial F^{(t)}_{ij}{}^z}{\partial r^{\alpha}_{i}} \\
\frac{\partial \widetilde{T}_{ij}^{z}}{\partial r_i^\alpha} & = - \left( \frac{\delta_{\alpha x}}{r_{ij}} - \frac{x_{ij}r_{ij}^\alpha}{r_{ij}^3} \right) F^{(t)}_{ij}{}^y - \left( \frac{x_{ij}}{r_{ij}} \right) \frac{\partial F^{(t)}_{ij}{}^y}{\partial r^{\alpha}_{i}} +
\left( \frac{\delta_{\alpha y}}{r_{ij}} - \frac{y_{ij}r_{ij}^\alpha}{r_{ij}^3} \right) F^{(t)}_{ij}{}^x + \left( \frac{y_{ij}}{r_{ij}} \right) \frac{\partial F^{(t)}_{ij}{}^x}{\partial r^{\alpha}_{i}}.
\end{split}
\end{equation}
where $\delta_{\alpha x}$ (similarly,  $\delta_{\alpha y}$ and  $\delta_{\alpha z}$) is the Kronecker delta, such that $\delta_{x x}=1$, $\delta_{y x}=0$ and $\delta_{z x}=0$, and
\begin{equation}
\begin{split}
 \frac{\partial \widetilde{T}_{ij}^{x}}{\partial \theta_i^\alpha} &= -\left[\left(\frac{y_{ij}}{r_{ij}} \right) \frac{\partial F^{(t)}_{ij}{}^z}{\partial \theta_i^\alpha} - \left(\frac{z_{ij}}{r_{ij}} \right) \frac{\partial F^{(t)}_{ij}{}^y}{\partial \theta_i^\alpha} \right]\\
\frac{\partial \widetilde{T}_{ij}^{y}}{\partial \theta_i^\alpha} &= -\left[\left(\frac{z_{ij}}{r_{ij}} \right) \frac{\partial F^{(t)}_{ij}{}^x}{\partial \theta_i^\alpha} - \left(\frac{x_{ij}}{r_{ij}} \right) \frac{\partial F^{(t)}_{ij}{}^z}{\partial \theta_i^\alpha} \right] \\
\frac{\partial \widetilde{T}_{ij}^{z}}{\partial \theta_i^\alpha} &= -\left[\left(\frac{x_{ij}}{r_{ij}} \right) \frac{\partial F^{(t)}_{ij}{}^y}{\partial \theta_i^\alpha} - \left(\frac{y_{ij}}{r_{ij}} \right) \frac{\partial F^{(t)}_{ij}{}^x}{\partial \theta_i^\alpha} \right].
\end{split} 
  \label{Eq:dTdtheta}
\end{equation}
The above two differential equations can be solved using~(\ref{Eq:dFsbdra}), and~(\ref{Eq:dFsbdtheta}). 

\subsection{Jacobian}
The dimension of Jacobian operator $\B J$ is force over length. To be consistent with the dimension we redefine the torque $T$ and rotational coordinate $\theta$ as
\begin{equation}
  {\tilde T}_i = \frac{T_i}{\sigma_i}, \ \ \ \textrm{and} \ \ \ {\tilde \theta}_i = \sigma_i\theta_i
  \label{Eq:rescaledT}
\end{equation}
In addition, the dynamic matrix has a contribution from the moment of inertia $I_i=I_0m_i\sigma_i^2$ as $\Delta{\B \omega}_i={\B T}_i/I_i\Delta t$. In our calculation, we assume that mass $m_i$ and $I_0$ both are one. The remaining contribution of $I_i$, i.e. $\sigma_i^2$, is taken care of by rescaling the torque and the angular displacement as ${\tilde T}_i$ and ${\tilde \theta_i}$~(\ref{Eq:rescaledT}). For $I_0\neq 1$, the contribution of $I_0$ can be correctly anticipated if we rewrite~(\ref{Eq:dsdt}) as below:
\begin{equation}
  \frac{\mathrm d {\B t}_{ij}}{\mathrm d t} = {\B v}_{ij} - {\B v}^n_{ij} + \frac{1}{I_0}\hat{r}_{ij}\times(\sigma_i{\B \omega}_i + \sigma_j{\B \omega}_j),
  \label{Eq:dsdtnew}
\end{equation}

$\B J$ essentially contains {\bf four} different derivatives:
\begin{itemize}
\item First type: Derivative of force with respect to the position of particles:
  \begin{equation}
    \label{Eq:Jabij}
    \begin{split}
      & A^{\alpha\beta}_{ij} = \sum_{k=0; k\neq j}^{N-1}\frac{\partial F^\beta_{kj}}{\partial r^\alpha_{i}} = \frac{\partial F^\beta_{ij}}{\partial r^\alpha_{i}}, \ \ \ \textrm{for}\ \ i\neq j \\
      & A^{\alpha\beta}_{ii} = \sum_{j=0; j\neq i}^{N-1}\frac{\partial F^\beta_{ji}}{\partial r^\alpha_{i}} = -\sum_{j=0; j\neq i}^{N-1} A^{\alpha\beta}_{ij},
    \end{split}
  \end{equation}
where $N$ is the total number of particles. $A^{\alpha\beta}_{ij}$ is symmetric if we change pairs, i.e.: $A^{\alpha\beta}_{ij}=A^{\alpha\beta}_{ji}$, however the symmetry is not guaranteed with the interchange of $\alpha$ and $\beta$.
\item Second type: Derivative of force with respect to rotational coordinate:
   \begin{equation}
    \label{Eq:Jbij}
    \begin{split}
      & C^{\alpha\beta}_{ij} = -\sum_{k=0; k\neq j}^{N-1}\frac{\partial F^\beta_{kj}}{\partial \tilde\theta^\alpha_{i}} = -\frac{\partial F^\beta_{ij}}{\partial \tilde\theta^\alpha_{i}}, \ \ \ \textrm{for}\ \ i\neq j \\
      & C^{\alpha\beta}_{ii} = -\sum_{j=0; j\neq i}^{N-1}\frac{\partial F^\beta_{ji}}{\partial \tilde\theta^\alpha_{i}} = -\sum_{j=0; j\neq i}^{N-1} C^{\alpha\beta}_{ij}.
    \end{split}
   \end{equation}
   The negative sign makes sure that in stable systems all the eigenvalues are positive. $C^{\alpha\beta}_{ij}$ is asymmetric: $C^{\alpha\beta}_{ij}=-C^{\alpha\beta}_{ji}$.
 \item Third type: Derivative of torque with respect to position:
   \begin{equation}
     \label{Eq:Jaij}
     \begin{split}
       & B^{\alpha\beta}_{ij} = \sum_{k=0; k\neq j}^{N-1}\frac{\partial {\tilde T}_{kj}^\beta}{\partial r^\alpha_{i}} = \frac{\partial {\tilde T}_j}{\partial r^\alpha_{i}}, \ \ \ \textrm{for}\ \ i\neq j \\
       & B^{\alpha\beta}_{ii} = \sum_{j=0; j\neq i}^{N-1}\frac{\partial {\tilde T}_{ji}^\beta}{\partial r^\alpha_{i}} = \sum_{j=0; j\neq i}^{N-1} B^{\alpha\beta}_{ij}.
     \end{split}
   \end{equation}
$B^{\alpha\beta}_{ij}$ is also asymmetric: $B^{\alpha\beta}_{ij}=-B^{\alpha\beta}_{ji}$.
 \item Fourth type: Derivative of torque with respect to rotational coordinate:
   \begin{equation}
     \label{Eq:Jij}
     \begin{split}
       & D^{\alpha\beta}_{ij} = -\sum_{k=0; k\neq j}^{N-1}\frac{\partial {\tilde T}_{kj}^\beta}{\partial \tilde\theta^\alpha_{i}} = -\frac{\partial {\tilde T}_j}{\partial \tilde\theta^\alpha_i}, \ \ \ \textrm{for}\ \ i\neq j \\
       & D^{\alpha\beta}_{ii} = -\sum_{j=0; j\neq i}^{N-1}\frac{\partial {\tilde T}_{ji}^\beta}{\partial \tilde\theta^\alpha_{i}} = \sum_{j=0; j\neq i}^{N-1} D^{\alpha\beta}_{ij}.
     \end{split}
   \end{equation}
   The negative sign makes sure that in stable systems all the eigenvalues are positive. $D^{\alpha\beta}_{ij}$ is symmetric: $D^{\alpha\beta}_{ij}=D^{\alpha\beta}_{ji}$.
\end{itemize}

\subsection{Arrangement of Jacobian matrix}
All the J-matrix elements are combined together with the
following arrangement:
\arraycolsep=1.4pt\def\arraystretch{1.8}
\begin{equation}
\textbf{J} =
\left[
\begin{array}{ccc|ccc}
\frac{{\partial}F_x}{{\partial}x} ~&~\frac{{\partial}F_y}{{\partial}x} ~&~\frac{{\partial}F_z}{{\partial}x} ~&~\frac{{\partial}{\tilde T}_x}{{\partial}x} ~&~\frac{{\partial}{\tilde T}_y}{{\partial}x} ~&~\frac{{\partial}{\tilde T}_z}{{\partial}x} \\
\frac{{\partial}F_x}{{\partial}y} ~&~\frac{{\partial}F_y}{{\partial}y} ~&~\frac{{\partial}F_z}{{\partial}y} ~&~\frac{{\partial}{\tilde T}_x}{{\partial}y} ~&~\frac{{\partial}{\tilde T}_y}{{\partial}y} ~&~\frac{{\partial}{\tilde T}_z}{{\partial}y} \\
\frac{{\partial}F_x}{{\partial}z} ~&~\frac{{\partial}F_y}{{\partial}z} ~&~\frac{{\partial}F_z}{{\partial}z} ~&~\frac{{\partial}{\tilde T}_x}{{\partial}z} ~&~\frac{{\partial}{\tilde T}_y}{{\partial}z} ~&~\frac{{\partial}{\tilde T}_z}{{\partial}z} \\
\hline
\frac{{\partial}F_x}{{\partial}{\tilde \theta}_x} ~&~\frac{{\partial}F_y}{{\partial}{\tilde \theta}_x} ~&~\frac{{\partial}F_z}{{\partial}{\tilde \theta}_x} ~&~\frac{{\partial}{\tilde T}_x}{{\partial}{\tilde \theta}_x} ~&~\frac{{\partial}{\tilde T}_y}{{\partial}{\tilde \theta}_x} ~&~\frac{{\partial}{\tilde T}_z}{{\partial}{\tilde \theta}_x}\\
\frac{{\partial}F_x}{{\partial}{\tilde \theta}_y} ~&~\frac{{\partial}F_y}{{\partial}{\tilde \theta}_y} ~&~\frac{{\partial}F_z}{{\partial}{\tilde \theta}_y} ~&~\frac{{\partial}{\tilde T}_x}{{\partial}{\tilde \theta}_y} ~&~\frac{{\partial}{\tilde T}_y}{{\partial}{\tilde \theta}_y} ~&~\frac{{\partial}{\tilde T}_z}{{\partial}{\tilde \theta}_y}\\
\frac{{\partial}F_x}{{\partial}{\tilde \theta}_z} ~&~\frac{{\partial}F_y}{{\partial}{\tilde \theta}_z} ~&~\frac{{\partial}F_z}{{\partial}{\tilde \theta}_z} ~&~\frac{{\partial}{\tilde T}_x}{{\partial}{\tilde \theta}_z} ~&~\frac{{\partial}{\tilde T}_y}{{\partial}{\tilde \theta}_z} ~&~\frac{{\partial}{\tilde T}_z}{{\partial}{\tilde \theta}_z}\\
\end{array}
\right]
=
\left[
\begin{array}{c|c}
   \textbf{A} & \textbf{B} \\
   \hline
   \textbf{C} & \textbf{D}
\end{array}
\right]
\end{equation}
Every element of the above J-matrix is expanded into $N{\times}N$ sub-elements corresponding to i-j particle pairs. The total size of the matrix for $D=3$ is therefore $(2D)N\times(2D)N$.

\end{widetext}
\newpage
\bibliography{ALL}
\end{document}